\documentclass[12pt,fleqn]{article}

\usepackage[latin1]{inputenc}    

\usepackage[intlimits]{amsmath}

\usepackage{amssymb}

\usepackage{graphicx}

\usepackage{epsfig}

\title{Production and Width of the Pentaquark $\Theta^+ (1540)$ in Strangeness Production in the 
Proton-Proton System \thanks{supported in part by the Forschungszentrum FZ Jülich (COSY)}
\thanks{preprint FAU-TP3-06/Nr. 11}}

\author{M. Dillig \thanks{email: mdillig@theorie3.physik.uni-erlangen.de} \\
Institute for Theoretical Physics III \\ 
University Erlangen-Nürnberg,  \\ Staudtstr. 7, D-91058 Erangen, Germany }

\date{}
\begin{document}
\begin{titlepage}

\maketitle

\begin{abstract}

We analyze recent experimental information on the excitation of the $\Theta^+(1540)$ pentaquark in the 
hadronic reaction $pp \rightarrow p\Sigma^+ K^0$. Upon describing the conventional production process - 
which serves as a normalization - in a meson exchange model, we estimate the resonant $\Theta$ induced 
cross section and the width of the pentaquark via K - exchange in a meson-baryon and quark-gluon model. 
From a comparison with experiment we extract information on the width of the $\Theta^+$, the $p \Sigma^+ K^0$ 
coupling constant in the meson exchange and the relative s or p wave function for a nucleon-kaon cluster 
in the $\Theta^+$.

%\end{abstract}

\hspace*{0.6cm}pentaquark $\Theta^+$.

\end{abstract}
\end{titlepage}
%%%%
\newpage
\setcounter{page}{2}
%{\bf 1. Introduction} \\
\section{Introduction}

It is well known that QuantumChromoDynamics QCD allows the existence of colorless exotic hadronic states, 
beyond the conventional $q \overline q$ or 3q  quark structure for mesons and baryons \cite{Eidel}. 
In fact there has been a long history in the search and investigation of such states. From a Fock expansion 
of hadrons

\begin{displaymath}
B> = \alpha |qqq > + \beta |qqq q\overline q > + ....
\end{displaymath}

the strong coupling to (on-shell) nucleon-meson final states for baryons above the nucleon ground 
state with typical decay widths of $\Gamma_B \sim 100 - 300$ MeV already points to strong $q \overline q$ 
components in baryons (\cite{Hofmannqq}\cite{Schott}). \\

In this respect the $\Theta^+(1540)$ pentaquark, predicted years ago in a soliton model \cite{Diakonov} 
and reported recently from the LEPS collaboration at SPring8 \cite{LEPS}, which is currently understood 
as a $uuu d \overline s$ or $udd u \overline s$ resonance in the $p K^0$ or $n K^+$ system, did not come as a surprise. 
However, very surprising is its very small width of $\Gamma_{\Theta} \leq 20 MeV << \Gamma_B$ for conventional 
baryon resonances.\\

Since the first report and the accumulating evidence from numerous labs on the $\Theta^+$ pentaquark 
%\cite{DIANA},\cite{CLAS},\cite{SAPHIR},\cite{Astrapen},\cite{CLASneu},\cite{HERMES},\cite{ZEUS},
%\cite{COSYpen},\cite{SVD},\cite{CLASneuneu} (for experimental surveys see ref. %\cite{Burkert},\cite{Schumacher}),
(\cite{DIANA} - \cite{Burkert}) (for a critical review of the experimental situation see ref. (\cite{Schumacher},\cite{Burkcrit}),
a flurry of theoretical ideas and investigations of the excitation and the structure of the pentaquark appeared 
in the literature 
%\cite{SibKN},\cite{CarlAnti},\cite{Liupgam},\cite{Jaffe},\cite{Jennings},\cite{Karliner},
%\cite{Lipkin},\cite{Koch},\cite{CarlsonPrag},\cite{JaffeJain},\cite{PInstant},\cite{CarNatural},\cite{},
%\cite{Nam32},\cite{Oh},\cite{Hossurvey},\cite{Zahed},\cite{Stech},\cite{Gal1},\cite{Riska},\cite{Bucella},
%\cite{},\cite{},\cite{};
(\cite{PolyakNN} - \cite{PInstant});
 a survey on Pentaquarks up to the year 2004 is given in ref. \cite{Penta2004}. Unfortunately, 
in practice a rigorous calculation of the strongly interacting 5 quark system is extremely difficult (compare, 
however, recent work from E. Hiyama et al. \cite{Toki}). Thus, guided by findings from nuclear 5-body systems 
$^5 He$ or $^5 _\Lambda He$ and accumulating evidence for quark-quark correlations in specific spin-isospin 
channels, various cluster models for the pentaquark have been proposed; among them, as widely accepted, a 
$(qqq) (q \overline q)$ meson - nucleon like component (\cite{Koch},\cite{Karlip}) or a $(qq) (qq) \overline q$ 
substructure (with scalar-isocalar diquarks \cite{Jaffe}). However, all of these models are still in a very 
rudimentary stage; presently even basic details about the quantum numbers of the $\Theta^+$ pentaquark, 
such as its parity, its spin and isospin, or quantitative details about its structure and width are still missing.\\

In this note we address various aspects of the $\Theta^+$, its excitation and its structure, combining 
simple model assumptions with experimental information. Explicitly we focus for the nucleon-nucleon system 
on strangeness production, i. e. on the exclusive reaction 
$pp \rightarrow p \Sigma^+ K^0$, performed recently from the COSY-TOF collaboration \cite{COSYpen}. 
From their data at a beam momentum of 2.95 GeV/c they extract the total nonresonant and resonant cross 
sections and their ratio

\begin{displaymath}
 R = \frac{\sigma_{NR}}{\sigma_R} \sim \frac{12.4 \mu b}{0.4 \mu b} \sim 30
\end{displaymath}

A confrontation of these data with theoretical models seems very promising, as comprehensive experimental (and theoretical) information on 
related strangeness production in the pp system, i. e. $pp \rightarrow p \Lambda K^+, p \Sigma^0 K^+$ total 
cross sections and angular distributions at various energies, allow to pin down details of the nonresonant 
$pp \rightarrow p \Sigma^+ K^0$ cross section. For our interpretation of the data and for an estimate 
of the excitation and the properties of the $\Theta^+$, we compare two different 
models: a formulation based on meson and baryon degrees of freedom and, as an alternative, a representation in terms of constituent quarks (and gluons). \\

Details we present in the following sections. In chapter 2 we derive a simple meson exchange model for the 
nonresonant $\Sigma^+ K^0$ cross section. The following chapter then contains a derivation of both the width 
of the pentaquark and its excitation in resonant $\Sigma^+ K^0$ production, comparing meson exchange with 
quark exchange. The various results and their discussion are presented in chapter 4, before summarizing 
and concluding in the final chapter. 
\vskip 1.5cm

\section{The nonresonant $\sigma_{NR} (pp \rightarrow p \Sigma^+ K^0)$ cross section}

In this section we sketch the basic steps to estimate the total cross section 
$\sigma_{NR} (pp \rightarrow p \Sigma^+ K^0)$ in a meson exchange model.\\

\subsection{The total cross section $\sigma_{NR}$ }

For the total cross section we define as kinematical variables in the overall CM 
system the momenta $\pm \bf k_p$ for the incoming protons and the momenta  ${\bf q-k/2} $, $\bf{-q-k/2}$ and ${\bf k}$ for the (outgoing) proton, $\Sigma$ and kaon, respectively; the magnetic spin 
 components are $\mu_1, \mu_2$ for the initial and $\mu_p, \mu_{\Sigma}$ for the final state. 
 Then the total cross section at an excess energy

\begin{displaymath}
Q = {\sqrt s} - (M_p + M_{\Sigma} + m_K) 
\end{displaymath}   

(with $\sqrt s$ as the total CM energy) is given as

%Formel 1
%
\begin{displaymath}  
\sigma_{NR} (Q)  =  \frac{1}{(2 \pi)^{5}} \; \frac{E_{p}}{2 k_{p}} \cdot \frac{1}{4} \; 
\sum_{\mu_{1} \mu_{2} \mu_{p} \mu_{\Sigma}}        
\int | T_{i \to f} ({\bf k}, {\bf q}, Q)|^{2} \delta (Q - T_{p} - T_{\Sigma} - T_{k}) \; 
\frac{d{\bf k} d {\bf q}}{2 \omega_{k} (k)}
\end{displaymath}  
with the kinetic energies $T_i$ for the different particles and the total energy of the kaon 
$\omega_K(k^2) = m_K + T_K(k^2)$.\\

\subsection{The $pp \rightarrow p \Sigma^+ K^0$ amplitude}

There are a variety of calculations for exclusive strangeness production to $\Lambda K^+$ and 
$\Sigma^0 K^+$ final states 
%\cite{Gasp},\cite{Sib},\cite{Shyam},\cite{Tsush},\cite{Edep},
(\cite{Tsush} - \cite{Edep}). As the experimental data point for the $\Sigma^+ K^0$ cross section is given at an excess energy 
of Q = 126 MeV, which is still very close to the threshold energy 
$E_{thr} = (M_p + M_{\Sigma} + m_K) - 2M_p \sim 750$ MeV, we rely on findings from meson exchange models for associate strangeness production. 
There in most cases, the total transition amplitude is factorized schematically as

\begin{displaymath}
|T_{i \rightarrow f}|^2 = |M_{FSI}|^2 \, . \, |M_{i \rightarrow f}|^2 \, . \, |M_{ISI}|^2
\end{displaymath}

into initial state interactions (ISI) in the pp  and final state interactions (FSI) in the $p \Sigma K$ 
system and the genuine production amplitude $M_{i \rightarrow f}$ \cite{HanRept} (for our purpose we 
integrate out final state interactions together with the production amplitude). Above the ISI reflects 
dominantly the coupling of the two protons to inelastic channels and can be approximately incorporated 
as an energy independent reduction of the cross section with typically $M_{ISI}^2 \sim 1/3$ \cite{HanNak}. 
Similarly, as the Q dependence of the total cross section over the restricted energy range of our interest, 
is dominated by phase space, $p \Sigma$ final state interactions are readily incorporated in an effective 
range - scattering length parametrization \cite{SibCass} (equivalent to a parametrization via Jost functions 
\cite{Sibphi}; K baryon final state interactions are expected to be weak and negligible for our estimate).\\

It is the current understandig that the production operator $M_{i \rightarrow f}$ is dominated by the 
excitation of various baryon resonances (Fig. 1), the s-wave resonance $N^*(1650, 1/2^-)$ and the p-wave 
resonances $N^*(1710, 1/2^+), N^*(1720, 1/2^+)$ (with the possible inclusion of the $\Delta(1960,3/2^+)$ 
resonance), which all show a strong decay into hyperon - kaon channels. As the sharing of the large momenta 
for the incoming protons is dominantly provided by single meson exchange, the production amplitude in 
leading order is given by
%
%Formel 2
%
\begin{eqnarray*}
M^{NR}_{i \to f} ({\bf k}_{p}, {\bf q}, {\bf k})  & = &  \sum_{i, k} 
\frac{L_{KN^{*}_{i} \Sigma } L_{\lambda_{k} p N^{*}_{i}}}{E^{*}_{N_{i}}(({\bf q} + {\bf k}/2)^{2}) 
+ E_{p} (({\bf q} - {\bf k}/2)^{2}) - \sqrt{s}}   \\
& & \frac{L_{\lambda_{k} pp}}{({\bf k}_{p} + {\bf k}/2 - {\bf q})^{2} + m^{2}_{\lambda} -  
\omega^{2}_{\lambda} ({\bf k}, {\bf q}) } \, ,
\end{eqnarray*}
where $L_{\lambda_k pp}$, $L_{\lambda_k NN^*}$ and $L_{KN^*\Sigma}$ denote the various meson-baryon vertex functions for the 
coupling of the mesons $\lambda_k$ (they include the corresponding operator, the coupling strength and a 
form factor). $E_p, E_{N^*}$ denote the total proton and (complex) energies of the baryon resonances; 
the different sums over i and k include all meson exchanges (i) and baryon resonances (k) excited. 
With these specifications the calculation of the total cross section is now straightforward. As we are 
ultimately only interested in a comparison with the ($\Theta^+$ dominated) induced cross section at a 
given Q value, we simplify our model building further in two steps. Thereby we note from kinematics 
for threshold energies $ Q \geq 100$ MeV of interest, that the variation of the meson propagators 
is, due to $(({\bf k_p} + {\bf k}/2 - {\bf q})^2 \sim k_p^2 >> q^2,k^2$, very small, so that the 
sum over the different meson exchanges and their propagators can be approximated by an overall constant, 
which simulates the exchange of an effective scalar meson with a mass of 550 MeV, as typical for 
$\sigma$-meson exchange. Then with the remaining static Lagrangians for the $\sigma$ and kaon coupling 
to the p and s wave baryons with positive (+) and negative (-) parity mentioned above,

\begin{displaymath}
L^+_{\sigma N N^*(N)} = g_{\sigma}^* (g_{\sigma})\\ 
L^+_{K N N^*(N)} =  \frac{f_K^* (f_K)}{m_K} {\bf \sigma q}
\end{displaymath}

and similarly

\begin{displaymath}
L^-_{\sigma N N^*} = \frac{f_{\sigma}^*}{m_{\sigma}} {\bf \sigma q} \\
L^-_{K N N^*(N)} = g_K^*
\end{displaymath}

the transition amplitude is given as
%
% Formel 3
\begin{displaymath}
M^{NR}_{i \to f}  ({\bf k}_{p}, {\bf q}, {\bf k})  \cong   \sum_{i}  
\frac{L^{\pm}_{KN^{*}_{i} \Sigma} L^{\pm}_{\sigma p N^{*}_{i}}}{E_{N^{*}_{i}} + E_{p} - \sqrt{s}} \; 
\frac{L_{\sigma pp}}{{\bf k}^{2}_{p} + m^{2}_{\sigma}}
\end{displaymath}
In a final step we even simplify the contributions from the baryon resonances. As in the sum over i we 
test for $Q \geq 100 MeV$ the smooth tails of the baryon resonances with a typical width of 150 MeV over 
a small energy interval, we sum up all resonance contributions from the various mesons in a single effective 
coupling as a scalar meson with an averaged baryon resonance. Combining these elements we end up with an 
effective production operator for the coupling to an s-wave resonance (corresponding to a dominant
excitation of the $N^*(1650)$ resonance)

\begin{displaymath}
M_{i \rightarrow f} ({\sqrt s},q,k) = \frac{g_K^* g_{\sigma}^*}{E_{N^*}+E_p-\sqrt s}
\, \frac{g_{\sigma} \, {\bf \sigma k_p}}{k_p^2 + m_{\sigma}^2},
\end{displaymath}   

which yields, upon averaging over initial and final state interactions, ultimately

\begin{displaymath}
|T^-_{i \rightarrow f}|^2 = (\frac{G}{M^2_N})^2
\end{displaymath}

with an overall effective coupling constant G; the nucleon mass $M_N$ is only introduced for dimensional 
reasons. A similar expression is obtained for the coupling to a p-wave resonance (corresponding to the 
dominant excitation of the $N^*(1710)$ and $N^*(1720)$), with the only basic difference that the kaon 
momentum enters explicitly 

\begin{displaymath}
|T^+_{i \rightarrow f}|^2 = (\frac{G'}{M^3_N})^2 k^2
\end{displaymath}

(the k dependence enters into the integration over phase space). When averaging over the resonances we feel 
on a safe ground, as the Q dependence of the ratio $\Lambda \slash \Sigma^0 \sim 2.5$  at Q = 126 MeV
has already dropped to the corresponding ratio at $Q \geq 300$ MeV, consequently resonance excitation contributes 
only in a very averaged smooth way, in contrast to the strong variation of the ratio 
$\Lambda \slash \Sigma^0 \sim 10 - 30$ for Q values below 60 MeV (\cite{Sewerin},\cite{Kow}) (which 
reflects selective contributions from the various resonances). Thus $\sigma_{NR}$ for strangeness production 
close to the $\Sigma$ threshold involves in the simplest approximation just a single parameter (times the 
integration over FSI and phase space), which can be fixed from the normalization of the cross section at Q=126 MeV.\\
\vskip 2.0cm

\section{Width of the pentaquark and the resonant $pp \rightarrow \Sigma^+ \Theta^+ \rightarrow p \Sigma^+ K^0$ 
cross section}

In the following we derive the width and the $\Theta^+$ induced cross section both in a meson exchange and 
constituent quark model. Thereby we assume throughout S = 1/2 and I = 0 for the spin and isospin quantum 
numbers of the $\Theta$ and compare the results for a negative or positive parity of the pentaquark.  \\

\subsection{Meson exchange}

The diagrammatic structure of the width and the resonant cross section in a meson exchange model are 
schematically given in Figs. 2(a),3(a); evidently both diagrams involve the $p \Theta^+ K^0$ vertex. 
Depending on the $\Theta^+$ parity, the vertex functions are given (similarly as in chapter 2) from a 
reduction of the corresponding relativistic Lagrangians

\begin{displaymath}
L_{N \Theta^+ K} = g_{ N\Theta^+ K}{\overline \Psi}_N (i\gamma_5,1) \Psi_{\Theta^+} \Phi_K
\end{displaymath}

with the operators $i\gamma_5$ and 1 corresponding to positive and negative $\Theta$ parity, respectively. 
The leading non relativistic counterparts are then given as%
\begin{displaymath}
L^-_{K N N^*(N)} = g_{ N \Theta^+ K} \\
L^+_{p \Theta^+ K} = \frac{g_{ N \Theta^+ K}}{2 M_N} {\bf \sigma k} 
\end{displaymath}

for the pK coupling to a $\Theta$ with positive and negative parity, respectively (for a $\Theta^+$ in rest).\\

With the standard formula
%Formel 4
\begin{displaymath}
\Gamma_{\Theta \to KN} = \frac{1}{(2 \pi)^{2}} \; \frac{1}{2} \; \sum_{\mu_{\Theta} \mu_{p}} \int  
| T_{\Theta \to KN} (k_{0}, \hat{{\bf k}}) |^{2} \frac{E_{N}(k)}{M_{\Theta}} \; k_{0} d \hat{{\bf k}}
\end{displaymath}
we then obtain from
%Formel 5
\begin{displaymath}
| T^{\pm}_{\Theta \to KN} ({\bf k}) |^{2} = \frac{1}{2} \sum_{\mu_{\Theta} \mu_p} |\langle 1/2 \mu_{p} | L^{\pm}_{N \Theta^{+}k} ({\bf k}) 
| 1/2 \mu_{\Theta} \rangle | ^{2} \, = \, 1 \, , 
\end{displaymath}
the total width of the pentaquark from the two open decay channels $p K^0$ and $n K^+$ as 

\begin{displaymath}
 \Gamma_{\Theta^+} = \Gamma_{pK^0}+\Gamma_{nK^+} = 2 \Gamma_{pK^0}
\end{displaymath}

(above we dropped small differences from the kinematics for the N and the kaon with different isospin projections). For the ratio, assuming the same strength for s and p wave coupling (which is certainly a very vague assumption), we end up with 

\begin{displaymath}
\Gamma_+ = \frac{1}{4} (\frac{k_0}{M})^2 \, \Gamma_- 
\end{displaymath}

For the corresponding final kaon (nucleon) momentum

\begin{eqnarray*}
k_0 & = &  \frac{1}{2M_{\Theta}} {\sqrt{(M_{\Theta}^2 + m_K^2 - M_N^2)^2 - 4 M_{\Theta}^2 m_K^2}} \\[2ex]
& &   \sim \;  {\sqrt {2 M_N m_K \, \frac{M_{\Theta}-M_N-m_K}{M_N+m_K}}} \, \sim \, 260 MeV
\end{eqnarray*}

this yields the well known suppression of more than one order of magnitude of the width of the pentaquark with positive compared to negative parity. Of course, the crucial input for the width is effective $N\Theta^+ K$ coupling constant; for its value and its interpretation we come back in the following chapter, when discussing our results.\\

The contribution of $\Theta^+$ excitation to the resonant cross section is conceptionally straight forward and follows similar lines as sketched above for the nonresonant cross section, though it is technically more complicated. For the calculation of the $\sigma_R$ we follow the reasoning from chapter 2, where in addition now the intermediate kaon rescatters on the second nucleon to excite the pentaquark (Fig. 2(a)). The competitive mechanism, where the kaon is first created directly at the $p\Sigma^+$ vertex (which would correspond to the direct 'single nucleon emission' in $\sigma_{NR}$) and rescatters subsequently and excites the $\Theta^+$ on the proton, is expected to be small, as there the sharing of the large proton momenta in the initial state has to be provided exclusively by a K exchange far off its energy shell, which also probes the $N\Theta K$ vertex at very large momentum transfers. In general, the direct 'one nucleon emission' is suppressed substantially for heavy meson production (already in pion production its contribution is about one order of magnitude smaller than the piece from meson rescattering \cite{HanRept}).\\

With these ingredients the transition amplitude $M_R$ is given for an $N^*$ and $\Theta^+$ with negative parity as
%
%Formel 6
\begin{eqnarray*}
M^{R}_{i \to f} ({\bf k}_{p}, {\bf q}, {\bf k})  & = &  \frac{g^{2}_{N \Theta^{+}K}}{(2 \pi)^{3}}  \; 
\frac{1}{E_{\Theta}(({\bf q} + {\bf k}/2)^{2}) + E_{\Sigma} (({\bf q} - {\bf k}/2)^{2} ) - \sqrt{s}}  \\[2ex]
& & \int \frac{M^{NR}_{i \to f} ({\bf k}_{p}, {\bf q}, {\bf k}; {\bf p})}
              {({\bf p} - {\bf k})^2 + m_K^2 - (E_{N^*}({\bf p} - {\bf q} + {\bf k}/2)^2 - E_{\Sigma} ({\bf q} + {\bf k}/2)^{2})}   \\[2ex]
& & \cdot \frac{d {\bf p}}{( E_{\Sigma} (({\bf q} + {\bf k}/2)^{2}) + \omega_{k} (({\bf p} - {\bf k}/2)^{2} ) 
+ E_{N} (({\bf p} - {\bf q} + {\bf k}/2)^2 - \sqrt{s} )}
\end{eqnarray*}
where we introduced the corresponding nonresonant amplitude $M^{NR}_{i \rightarrow f}$ from chapter 2 together with K exchange in the loop integration over $d\bf p$; the final state is the characterized by the energy denominator for the $\Theta$ excitation and its subsequent decay into the KN final state. Again, as for the width of the $\Theta^+$ the transition from s to p wave coupling basically introduces an additional momentum squared in the loop from K exchange and an additional p-wave $N \Theta^+ K$ vertex from the coupling to the final state.\\

The evaluation of the loop for the rescattering contribution is in principle straightforward (both upon introducing 
form factors or in dimensional regularization, after subtracting the singular piece, the integral is finite), though in practice very cumbersome, as it involves a detailed parametrization of the nonresonant amplitude for associate strangeness production.
Here we proceed differently: as momentum sharing has been already provided from the scalar exchange in the 
initial amplitude and as the influence of final state interactions on the momentum and energy sharing is 
expected to be very weak, the nucleon and the kaon in the loop will propagate close to their mass shells.
Consequently, the by far dominant piece from the loop integration will result from the kinematics as given in the KN 
final state. Thus we restrict both the K and the N in the loop to their mass shell, keeping for the kaon propagator from the standard relation
\begin{displaymath}
\frac{1}{q^2-m^2-i\epsilon} = {\bf P}\frac{1}{q^2 - m^2} + i \pi \delta (q^2 - m^2)
\end{displaymath}  
only the imaginary component, yielding finally
the simple momentum dependent (complex) factor $-i \frac{k}{2 \pi}$ for the loop integration. \\

The remaining steps are now straightforward. Upon summing over the magnetic quantum number of the $\Theta^+$, we obtain for the squared transition amplitude
%Formel 7
\begin{eqnarray*}
| M^{NR}_{i \to f} (\sqrt{s}, {\bf q}, {\bf k} ) |^{2} & = &  |A (\sqrt{s}, {\bf q}, {\bf k}) |^{2}  \\
& & \cdot \left \{ 
\begin{array}{l}  g^{4}_{N \Theta K} 
\\ \left ( \frac{g^{2}_{N \Theta K}}{2 M_{N}} \right ) ^{2} (k^{2} + \lambda^{2} q^{2} - 2 \lambda 
{\bf k} {\bf q})     \end{array}  \right.   \text{for} \; 
\begin{array}{l} \Theta_{\pi = - 1} \\ \Theta_{\pi = + 1} 
\end{array}
\end{eqnarray*}
with
\begin{displaymath}
\big | A (\sqrt{s}, {\bf q}, {\bf k}) \big |^{2} = \frac{m_{k}Q}{(4 \pi)^{2}} \; 
\frac{1}{(E _{\Theta} (({\bf q} + {\bf k}/2)^{2}) + E_{\Sigma}(({\bf q} - {\bf k}/2)^{2} - 
\sqrt{s})^{2} + \Gamma^{2}_{\Theta} (({\bf q} + {\bf k}/2)^{2} 4 ) }
\end{displaymath}
for a $\Theta^+$ with negative or positive parity. In this last relation we kept for the p - wave $N \Theta^+ K$ vertex
%Formel 8
\begin{displaymath}
L_{N \Theta^{+} K} = i g_{N \Theta^{+} K} \langle U_{N} ({\bf q} - {\bf k}/2) | \gamma_{5} | \; 
U_{\Theta} ({\bf q} + {\bf k}/2) \rangle
\end{displaymath}
both the static and the leading nonstatic term

\begin{displaymath}
L_{N \Theta^+ K} = i\,  \frac{g_{N \Theta^+ K}}{2 M_N} {\bf \sigma} ({\bf k} - \lambda {\bf q})
\end{displaymath}

with $\lambda \sim 2/5$ for $M_{\Theta^+} \sim M_N + m_K \sim \frac{3}{2} M_N$.\\

The result above already reflects a consistency relation of the production amplitude and the width of the 
$\Theta^+$. From the on-shell assumption for the loop integration and upon dropping the small variation of 
the momentum dependence of the width of the pentaquark in the rescattering process, the width is strictly 
related (at the same kinematics) to the production amplitude. Thus $\sigma_R$ and $\Gamma_{\Theta^+}$ are 
in this limit rigorously tied together; comparison with the experimental nonresonant and resonant cross section should therefore set bounds for the excitation  and the width of the  
$\Gamma_{\Theta^+}$, or equivalently, for the coupling strength $N\Theta^+ K$ as the only parameter in our 
estimate.
\vskip 1.0cm

\subsection{Constituent quark exchange}

As an alternative to meson exchange we calculate the width and excitation of the pentaquark in a 
constituent quark model. The diagrammatic approach as sketched in Fig. 2(b), 3(b) remains the same as for
 meson exchange, except with the difference that the $N \Theta^+ K$ vertices are replaced by the corresponding 
 quark substructures. Here the nucleon and the K meson are represented as 3q and $q \overline q$ systems. 
 For the $\Theta^+$ details of its substructure are presently unknown; among the different models on the 
 market we focus here on a (3+2) nucleon-kaon clustering of the $4q \overline q$ of the $\Theta^+$ (adding 
 in our discussion a sketchy result for a diquark-diquark-$\overline q$ pentaquark structure as suggested 
 in ref. \cite{Jaffe})\\

Without knowing quantitative details  and for an analytical evaluation of the various matrix elements, we choose for the interacting quarks an harmonic oscillator basis, following the standard assumption that 
the 3 quarks in the nucleon and the $q \overline q$ pair in the kaon are in relative s states, whereas 
for the $\Theta^+$ we allow for a relative s or p wave between the nucleon and kaon cluster (to account 
for negative and positive parity of the $\Theta$). Then in going over to standard Jacobi coordinates the 
various wave functions are given as
%Formel 9
\begin{eqnarray*}
| K \rangle & = & R_{os} (r) |\chi^{S}_{00} \chi^{F}_{1/2 \tau_K} \chi^{C}_{00} \rangle  \, ; \\[2ex]
| N \rangle & = & R_{os} (r) R_{os} (\rho) |\chi^{S}_{1/2 \mu_{N}} \chi^{F}_{1/2 \tau_{N}} \chi^{C}_{00}\rangle
\end{eqnarray*}
and 
\begin{eqnarray*}
| \Theta \rangle & = &  | [NK] \rangle   \\
& = & (R_{os} (r) R_{os} (\rho))_{N} R_{0s} ({\lambda})_{0s} 
R_{L} (R) [Y_{L} (\hat{{\bf R}}) 1/2]1/2 \mu _{\Theta}  \\[1ex]
& & |[ \chi^{N}_{1/2} \chi^{K}_{1/2} ] ^{F}_{00} [\chi_{00}^N \chi^{K}_{00} ]^{C}_{00}\; \rangle
\end{eqnarray*}
with
\begin{displaymath}
R_{os} (r) = N_{os} \; e^{- \frac{r^{2}}{2 a^{2}_{i}}} 
\end{displaymath}
and
\begin{displaymath}
R_{L} (R) = N_{oL} (R/a_{R})^{L} \; e^{- \frac{R^{2}}{2 a^{2}_{R}}}
\end{displaymath}
for the kaon, the nucleon and the $\Theta^+$, respectively, where the oscillator parameters $a_i$ 
reflect the size of the particles. Above the spin (S), flavor (I) and color functions (C) are indicated 
implicitly (as they drop out in the overlap matrix elements between the KN clusters in the $\Theta^+$ 
and in the KN final state); the relative coordinate for the pentaquark involves the coupling of the 
orbital angular momentum L to the spin of nucleon cluster in the pentaquark (of course, for a negative 
parity state with L=0 the spin-orbital coupling factorizes); $N_{0s}, N_{0L}$ are appropriate normalization 
constants. It should be noted that the size parameters for the nucleon and kaon cluster in the pentaquark 
are presumably different from those in the free particles, though quantitative details are presently unknown. \\

With these prerequisits the width of the pentaquark is readily calculated along similar lines as for the 
meson exchange models. There are two basic differences. For the KN coupling to the $\Theta^+$ the KN$\Theta$ 
coupling constant is replaced by the corresponding overlap matrix element, schematically given as $<KN|\Theta>$. 
In addition, to estimate the coupling strength, we have to account for the transition from the free KN to 
the KN cluster in the pentaquark. As in nuclear physics this transition strength is proportional to the corresponding 
spectroscopic factor (\cite{CarNatural} \cite{Hossurvey}).

\begin{displaymath}
|S_{\Theta (NK)}|^2 = |<KN |\Theta>|^2 = (\sqrt \frac{5}{96})^2
\end{displaymath}\\

as the probability for the finding the KN configuration in the pentaquark.\\

With these ingredients $\Gamma_{\Theta^+}$ is readily obtained, following the definition of the width from 
the previous section and the explicit transition matrix element
%Formel 10
\begin{eqnarray*}
T^{\pm}_{\Theta \to NK} ({\bf k}) & = &  S_{\Theta (KN)}  
\langle \phi^{\prime}_{N} (r, \rho) \phi^{\prime}_{K} (\lambda) | \phi_{N} (r, \rho) \phi_{K} (\lambda) \rangle \\[1ex]
& & \langle e^{i{\bf k R}}, 1/2 \mu | R_{L}(R) [Y_{L} ({\bf \hat{R}}) 1/2 ] 1/2 \mu_{\Theta} \rangle 
\end{eqnarray*}
where - neglecting KN final state interactions - the cluster wave function is given by a plane wave 
(with $\bf k$ being the momentum of the outgoing kaon and proton; the flavor and color matrix elements yield unity). 
Upon integrating over the relative coordinates we find for the overlap integral
%Formel 11
\begin{eqnarray*}
I_{r \rho \lambda} & = &  \langle \phi^{\prime}_{N} (r, \rho) \phi^{\prime}_{k} (\lambda) | \phi_{N} (r, \rho) 
\phi_{k} (\lambda) \rangle  \\[2ex]
& = & \left ( 2 \; \frac{a_{r}a^{\prime}_{r}}{a^{2}_{r} + a^{2}_{r^\prime}} \right )^{3/2} \; %%
\left ( 2 \; \frac{a_{\rho}a^{\prime}_{\rho}}{a^{2}_{\rho} + a^{12}_{\rho}} \right )^{3/2} \; 
\left ( 2 \; \frac{a_{k} a^{\prime}_{k}}{a^{2}_{k} + a^{12}_{k}} \right )^{3/2} 
\end{eqnarray*}
which reduces to unity for identical size parameters for the nucleon and the kaon in the final state 
and in the $\Theta^+$. Similarly we find for the Fourier transform of the (normalized) relative cluster 
wave function
%Formel 12
\begin{displaymath}
I_{L} ({\bf k}) = N_{L} \int R^{L} Y_{LM} (\hat{{\bf R}}) \; 
e^{- \frac{R^{2}}{2 a^{2}_{R}} + i {\bf k} {\bf R}} d {\bf R}
\end{displaymath}
with
\begin{eqnarray*}
I_{0} ({\bf k}) & = &  \frac{2  a^{3/2}_{R}} {\pi^{1/4}} \; e^{- \frac{a^{2}_{k} k^{2}}{2}} 
Y_{00} (\hat{{\bf k}})  \\[2ex]
I_{1} ({\bf k}) & = & i  \frac{\sqrt {\frac{8}{3}} a^{3/2}_{R}}{\pi^{1/4}} (a_{R} k) \; 
e^{- \frac{a^{2}_{R}k^{2}}{2}} \; Y_{1M} (\hat{{\bf k}}) 
\end{eqnarray*}
for s and p states, respectively. All pieces are put together after decoupling orbital angular momentum 
and spin and summing over the external spin quantum numbers. Then, upon squaring the amplitude as above, 
%Formel 13
we find explicitly
%Formel 14
\begin{displaymath}  
\Gamma^{\pm}  =   \frac{64}{3 \pi} \; \frac{k_{0}E_{N}(k_{0}^{2})}{M_{\Theta}} \; S^{2}_{\Theta (kN)}
I^{2}_{r \rho \lambda} \big ( 
N_{L} \big )(a_{R}k_{0})^{2 L} \; e^{- a^{2}_{R} k^{2}_{0}}
\end{displaymath}
with
\begin{displaymath}
N_{0} = 1, \quad N_{1} = \frac{1}{3} \, .
\end{displaymath}
 for a pentaquark with negative and positive parity (above we included the factor $\frac{1}{2}$ for the p state from the Moshinksy transform from single particle to Jacobi coordinates). Comparing with the meson exchange estimate we find, as the flavor 
 and color matrix elements yield unity for the $\Theta^+$ width, that the $g_{N\Theta K}$ coupling constant 
 is basically replaced by the spectroscopic factor times the corresponding Fourier transform of 
 relative KN cluster wave function in the pentaquark (assuming that the overlap integrals from initial and final state for the internal K and N coordinates equals 1, which certainly provides an upper limit for the following estimates).\\

The resonant cross section $\sigma_R$ follows similar lines as the calculation of the width. Only two 
technical details are worth mentioning. As here the final state involves in addition the $\Sigma^+$ in 
a 3 particle continuum, the K and the nucleon are no more emitted back to back, but with the relative 
momentum ${\bf Q} = {\bf q} + {\bf k}/2$; thus for a practical evaluation we 
introduce $\bf Q$ as the variable in the Fourier transform of the KN cluster wave function (to avoid mixed 
momenta in the angular momentum function $Y_1({\bf \hat Q} = \hat {\bf {q+k/2}})$) and change 
correspondingly the phase space integration. Furthermore, the summation over magnetic quantum 
numbers is more complicated due to the product of 2 angular momentum functions already in the 
transition amplitude. For a compact result, we decouple spin and orbital angular momentum, schematically,
%Formel  15
\begin{eqnarray*}
\langle NK | \Theta \rangle \langle \Theta | KN \rangle  & \sim & I^{2}_{L} (Q) \cdot 
I^{2}_{r \rho \lambda} ( k, Q)  \\[1ex]
& &  \sum_{\mu_{\Theta}\mu^{\prime}_{N}} \langle L M 1/2 \mu | 1/2 \mu_{\Theta} \rangle 
\langle LM^{\prime} 1/2 \mu _{p} |
1/2 \mu_{\Theta} \rangle Y_{LM} (\hat{{\bf Q}}) Y_{LM^{\prime}}^{*} (\hat{{\bf Q}}) \, ,
\end{eqnarray*}
combine the angular momentum functions and recombine the m-dependent product to 3-j and 6-j symbols as
%Formel 16
\begin{eqnarray*}
& & \sum_{M_{\Theta}MM^{\prime}} (-1)^{M^{\prime}}  \left ( 
\begin{array}{ccc} L & 1/2 & 1/2  \\ M & \mu &  \mu_{\Theta}  \end{array}           \right ) \left (
\begin{array}{ccc} L & 1/2 & 1/2 \\ M^{\prime} & \mu_{p} & - \mu_{\Theta}   \end{array}  \right ) \left (
\begin{array}{ccc} L & L & J \\ M & -M^{\prime} & M_{J}  \end{array}               \right )   \\[2ex]
& & \hspace*{2.2cm}   = (- 1)^{1/2 + \mu} \left ( 
\begin{array}{ccc} J & 1/2 & 1/2 \\ M_{J} & \mu_{p} & - \mu  \end{array}   \right )  \left \{
\begin{array}{ccc} J & 1/2 & 1/2 \\ 1/2 & L & L  \end{array}     \right \} \, 
\end{eqnarray*}
Upon squaring the full amplitude, summing over the spin projections of the nucleon and combining the 
angular momentum functions to the spherical $Y_{00}$, we end up finally with
%Formel 17
\begin{eqnarray*}
| T^{R} ({\bf Q}, {\bf k}) |^{2} & = & \frac{1}{\pi} A (\sqrt{s}, {\bf Q}, {\bf k}) |^{2} I^{2} _{L} (Q) 
I^{2}_{r \rho \lambda} (k) \left ( \frac{d \hat{{\bf Q}}}{4 \pi} \right )  \\[2ex]
& & \sum_{J} (\hat{L}^{2} \hat{J}^{2})^2    \left ( 
\begin{array}{ccc} L & L & J \\ 0 & 0 & 0 \end{array}  \right )^{2} \;  \left \{ 
\begin{array}{ccc} J & 1/2 & 1/2 \\ 1/2 & L & L  \end{array}    \right \}^{2}\; \delta_{J0}
\end{eqnarray*}
Evidently, the result from quark exchange can again be directly related to the corresponding piece on the meson exchange model: except of overall constants  we identify the corresponding 
$g_{N\Theta^+K}$ coupling constant as given by the overlap of the relative cluster wave functions in the $\Theta^+$ (with the size parameter $a_R$ and the final KN state), i. e.
%Formel 18
\begin{displaymath}
g^{2}_{N \Theta K} \sim (a_R Q)^{2L} exp(-a_R Q)^2
\end{displaymath}
for a $\Theta^+$ with negative $\pi = -1$ (L=0) or positive $\pi = +1$ (L=1) parity, respectively. Compared to the width of a $\Theta^+$ in rest, the
integral over $\Theta^+$ now depends explicitly on the momentum 
of the $\Theta$.  

\vskip 1.5cm

{\bf 4. Results and Discussion}\\

To arrive at specific conclusions on the nature the pentaquark, we have to specify our input and 
find support for the reasoning in our estimate.\\

As a first confrontation with experiment we confirm that our model building for the nonresonant 
$pp \rightarrow N\Sigma K$ cross section $\sigma_{NR}$ is qualitatively reliable. As for $\Sigma^+$ production the only published data point is at Q=126 MeV \cite{COSYpen}, we compare with data for $\Sigma^0$ production up to $Q \leq $ 60 MeV (\cite{Sewerin},\cite{Kow}) and find qualitative agreement, independent of subtle details of the model (provided the baryon resonances quoted in chapter 2 are included). This is 
in line with similar findings from earlier meson exchange calculations and also from a systematic comparison with 
$\Lambda K^+$ production, where detailed data both for the total cross section as well as angular distributions 
at different energies are available, single meson rescattering (including moderate final state interactions) 
reproduce the characteristic cross sections (note that lacking quantitative information on the proton-proton ISI 
prohibits absolute predictions; normalization has to be fixed from an experimental data point). \\

Backed by these findings we address the resonant cross section $\sigma_R$ and the $\Theta^+$ width. 
In the meson exchange model the only parameter in the calculation is the $N \Theta^+ K$ coupling constant. 
Unfortunately here little is known quantitatively. We find that the experimental ratio, quoted in 
the introduction,  $\sigma_{NR}/\sigma_R \sim 30$, is qualitatively reproduced for a $\Theta^+$ with positive parity and coupling constants 
$2.4 \leq g_{N\Theta^+K} \leq 4.5$ (Fig. 4(a)), which restricts the width of the $\Theta^+$ to $\leq$ 15 MeV and rules out a pentaquark with negative parity (Fig. 5(a)). However, the very restricted experimental information presently available, together with the very smooth dependence of the ratio $\sigma_{NR}/\sigma_R$ on $g_{N\Theta^+ K}$, prohibits for the moment quantitative conclusions.  
Here quantitative insight can come only from further experimental constraints (presently as well as from more theoretical 
information on the pentaquark structure).\\

To add one remark: from discussions in the literature it is not fully clear, how to define the 
$N \Theta^+ K$ coupling constant. The value quoted above contradicts estimates based on naive dimensional 
analysis, which yields $g_{N \Theta^+K}^{ND} \sim 4 \pi$ \cite{Carlsonnaive} and is thus larger by more 
than one order of magnitude for the squared coupling constant quoted above \cite{Liupgam}. As a remedy 
and to reconcile both findings it is argued to multiply is the result from dimensional analysis with the 
spectroscopic KN factor of the $\Theta^+$, yielding

\begin{displaymath}
g_{N \Theta K} \sim g_{N \Theta K}^{ND} \sqrt \frac{5}{96} \sim 3
\end{displaymath}

It is not clear, if such an identification makes sense, as in the mesonic picture the K meson, when 
coupling to the $N \Theta$ vertex, does not resolve the internal quark structure, so the rescaling of 
the $N\Theta K$ coupling constant above seems somewhat misleading.\\

 The drastic difference between the two coupling 
constants quoted above is not surprising. As well known from meson exchange calculations for the excitation 
of baryon resonances, such as the excitation of the $N^*(1535)$ resonance in $\eta$ production show in part striking differences to
estimates within naive dimensional analysis \cite{Dilleta}: the coupling constants there, as derived from 
various sources, are not natural, i. e. of the order of 1 in appropriate units. In addition, for any meson 
exchange calculation the coupling constants are extracted directly from the corresponding partial width 
of a resonance, without introducing additional spectroscopic factors. This holds in particular also for 
the $N^*(1710)$, being classified as the $\Theta(1540)$ as a member of the same $4q \overline q$ anti decuplet,
which would then for consistency suggest also the introduction of an additional spectroscopic factor, 
opposite to the standard extraction of meson $\lambda N N^*(1710)$ coupling constants from the partial 
decay width into the corresponding mesonic channel. Of course, these remarks are preliminary, still 
further insight has to be gained for a more definitive answer.\\

Turning to the quark model estimate of the $\Theta^+$ properties, the main findings from meson exchange 
are qualitatively confirmed for positive parity: the experimental ration $\sigma_{NR}/\sigma_R \sim 30$ is reproduced within the experimental limits for a width parameter $0.4 \leq a_R \leq 0.7$ in the KN cluster wave function (Fig. 4(b)), corresponding to a width of $\Gamma_{\Theta^+} \leq $ 8 MeV (Fig. 5(b)); again the excitation of a pentaquark with negative parity is ruled out from the data. However, already a very crude estimate shows, that the cluster structure of the $\Theta^+$ is crucial for a quantitative understanding. For example, estimating the width of the pentaquark with positive parity, but assuming as (scalar) diquark-diquark-quark 
substructure of the $\Theta^+$ (\cite{CarNatural}\cite{Hossurvey}), yields from the reduced KN spectroscopic factor (simply a consequence of the recoupling of quark lines from the $\Theta^+$ to the KN final state)

\begin{displaymath}
\Gamma_{JW} \sim \frac{96}{576} \Gamma_{KN} \leq 1 MeV
\end{displaymath}

 and thus a very small $\Theta^+$ width. However, further conclusions 
from the COSY data on the structure and the width of the pentaquark $\Theta^+(1540)$ are at the moment far too preliminary, as again the limited experimental information prevents a quantitative extraction of details of the KN cluster wave function, characterized by the size parameter $a_R$. \\

Valuating our findings, they map out at least on a qualitative level characteristic features of the $\Theta^+$. As expected from other investigations in the literature, they favor a pentaquark with positive parity (i. e. $L_{KN} =1$) and a width smaller than 10 MeV. A much narrower $\Theta^+$ may be obtained for a weak KN coupling constant $g_{N\Theta^+ K} \leq 1$ in the meson exchange picture. In the quark model the picture is less clear. A small $\Gamma_{\Theta^+}$ is obtained both for a very hard and a very soft pentaquark: for small $a_R \leq 0.4 fm$ the width scales with $\Gamma_{\Theta^+} \sim a_R^2$, while for very large $a_R > 1 fm$, the Gaussian part of the wave function takes over with $\Gamma_{\Theta^+} \sim \exp {(- a_R^2 k_0^2)}$ and enforces a small width of the pentaquark. In addition, the cluster structure of the $\Theta^+$ plays a crucial role, favoring for a $\Theta^+$ with a very small width a (2+2+1) diquark-diquark-$\overline q$ compared to a (3+2) KN clustering of the pentaquark. However, this estimate is very preliminary and needs further theoretical support.\\

\vskip 0.3cm

{\bf 5. Summary and Outlook}\\

In this note we investigate properties of the recently discovered pentaquark $\Theta^+(1540)$ in context 
with recent data on near threshold strangeness production in $pp \rightarrow p\Sigma K$, where for the 
$\Sigma^+ K^0$ final state a nonresonant and a resonant ($\Theta^+$ - induced) cross section at an excess 
energy of Q = 126 MeV has been extracted. Based on current ideas of nonresonant strangeness production 
in meson-baryon and constituent quark-gluon models, $\sigma_R$ and $\Gamma_{\Theta}$ are estimated. We 
find for coupling constants and size parameters of the interacting hadrons a typical width of 
less than 10 MeV for a pentaquark with positive parity (corresponding to a relative L=1 angular momentum in 
the nucleon - kaon cluster function in the $\Theta^+$); for a diquark clustering of the pentaquark its width is reduced by more than a factor 5 close to 1 MeV or less. Still, further quantitative conclusions, 
also on the structure of pentaquark, are presently absent.\\

Of course, the limitations of our estimate above are evident. Most serious is here the very approximate 
evaluation of the loop integration for the rescattering of the kaon: a more rigorous integration of the 
loop will certainly add real contributions from the rescattering diagram and thus lead to interference 
terms between the background (i. e. the nonresonant) and the resonant formation of the $p \Sigma^+ K^0$ final state. Unfortunately, the actual experimental information from the COSY - TOF experiment prohibits from conclusions on the pentaquark.\\

Beyond that an investigation of different aspects of the internal structure of the pentaquark, 
and in general of exotic 4q - $\overline q$ systems, is absolutely necessary. From the theoretical side this 
will include different assumptions on the cluster structure of the pentaquark - including 3+2, 4+1 and 2+2+1 
configurations - up to recent findings from a full blown 5-body Gaussian calculation of such systems. 
On the same level, a more detailed investigation of the non strange baryon resonance $N^*(1710)$ as a member of the anti decuplet and its quark content, beyond the standard 3q structure, has to be 
addressed (also in context with experimental information, such as its decays into specific meson-baryon 
channels). Other topics of interest are a different spin-isospin content of the $\Theta^+$ - for example 
a total spin S=3/2 would allow the admixture of higher angular momentum components \cite{Nam32} - or a more 
realistic investigation of the width of the $\Theta^+$ on the quark level. Already the spectroscopic decomposition of the (3+2) 
clustering of the pentaquark yields (\cite{CarNatural}\cite{Hossurvey})

\begin{displaymath}
|\Theta^+> \, = \, \frac{1}{2} |KN> + \sqrt \frac{1}{12} |K^*N> + 
\sqrt \frac{2}{3} |(q\overline q)_8 (qqq)_8> \, ,
\end{displaymath}

which indicates a sizeable content of $K^*N$ and color octet components in the $\Theta^+$.
This will modify the $\Theta^+$ width, for example from the transition of color octet configurations, 
following the appropriate interchange of quark lines or dynamically induced transitions from one and 
2 gluon exchange. Clearly, a detailed answer can only come from a microscopic calculation, which includes the dynamics of quarks and gluons in the decay amplitude. Finally, an extension of our theoretical basis 
to other reactions, where $\Theta^+$ production has been seen, might be interesting. For example, if our 
basic assumption that the intermediate (rescattering) kaon and the nucleon are close to their mass shell 
in exciting the $\Theta^+$ is correct, then a rather direct relation to KN scattering (\cite{SibKN},
\cite{Gal1}) via amplitudes on the mass shell would be possible. In fact, a further analysis of KN scattering seems very promising, as there in the CM system the $\Theta^+$ is produced in rest at a fixed energy (due to the two-body kinematics). In a sequential excitation, such as in pp or $\gamma p$ processes, the $\Theta^+$ enters with non vanishing momentum due to the 3 or 4 body kinematics; integrating then over phase space washes out the pole and the width of the pentaquark.\\

Of course, as frequently stressed above, the main impetus has to come from experiment. Focussing on forthcoming COSY activities \cite{EyrichGen}, one crucial 
experiment will be the determination of the parity of the pentaquark in strangeness production at COSY 
 with doubly polarized protons, by measuring the asymmetry coefficient $A_{yy}$ 
(\cite{TomPar} - \cite{HosPar}). Further information will come from the additional 
strange channel $pp \rightarrow n \Sigma^+ K^+$ (\cite{EyrichErice} - \cite{Karsch}), which in 
addition is interesting from its selectivity 
in the nonresonant production process (where - for small contributions from $K^0 - K^+$ charge exchange 
in the final state interaction, only the $\Delta^{++}(1960)$ should yield a substantial contribution to 
the cross section). Finally, a reanalysis of the $p \Sigma^+ K^0$ data from COSY - TOF, which is presently 
performed \cite{Eyrichpriv}, will set more stringent limits on the underlying experimental nonresonant and resonant 
cross sections.

%as charge exchange contributions from $n K^+ \leftrightarrow p K^0$ are expected#%ed to be small \footnote{A crude geometrical estimate for the $n K^+ \rightarrow% p K^0$ charge exchange amplitude is given by simple recoupling of spin, flavor% and color component of the leading quark exchange contribution. This gives for% the ratio
%Formel 19
%
%\begin{displaymath}
%\frac{R_{CE\chi}}{R_{NoCE\chi}}  =   \frac{T_{k^{+}n} \to k^{0} p}{T_{kN \to KN}} =  \left [
%\begin{array}{ccc} S & 1/2 & 1/2 \\ 1/2 & 1/2 & 0 \\ 1/2 & 0 & 1/2   \end{array}    \right ] \; \left [
%\begin{array}{ccc} T & 1/2 & 1/2 \\ 1/2 & 0 & 1/2 \\ 1/2 & 1/2 & 1/2  \end{array}  \right ] 
%\cdot \frac{1}{18} \; \epsilon_{ijk} \epsilon_{ije} \delta_{em} \delta_{km}
%\end{displaymath}
%
%and thus

%\begin{displaymath}

%R_{CEX} = (\frac{1}{6})_S; (\frac{1}{12})_A) R_{No CEX} 

%\end{displaymath}

%for a scalar S=T=0 or axial S=T=1 diquark in the nucleon. Thus the correspondin%g cross sections yield $\sigma_{CEX}\slash \sigma_{No CEX} << 1 \slash 10$ and %thus a suppression of KN charge exchange by more than one order of magnitude}.

\newpage

\newpage

\begin{figure}[h!]
\begin{center}
\includegraphics[width=12.0cm,height=10.0cm,angle=0]{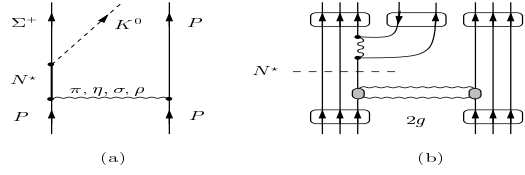}
\caption{Schematical representation of associate (nonresonant) strangeness production $pp \rightarrow p\Sigma^+K^0$ in a meson exchange (a) and a constituent quark model (b) ($N^*$ indicate intermediate baryon resonances; see text).}
\end{center}
\end{figure}

\begin{figure}[h!]
\begin{center}
\includegraphics[width=12.0cm,height=10.0cm,angle=0]{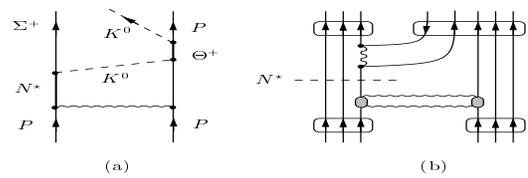}
\caption{As Fig. 1, however, for the (resonant) cross section, including the excitation of the $\Theta^+$ pentaquark.}
\end{center}
\end{figure}

\begin{figure}[h!]
\begin{center}
\includegraphics[width=12.0cm,height=10.0cm,angle=0]{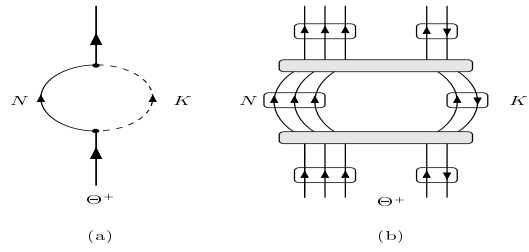}
\caption{Width of the $\Theta^+(1540)$ in a meson exchange (a) and a quark model (b) (assuming a (2+3) KN cluster structure for the pentaquark).}
\end{center}
\end{figure}

\begin{figure}[h!]
\begin{center}
\includegraphics[width=12.0cm,height=10.0cm,angle=0]{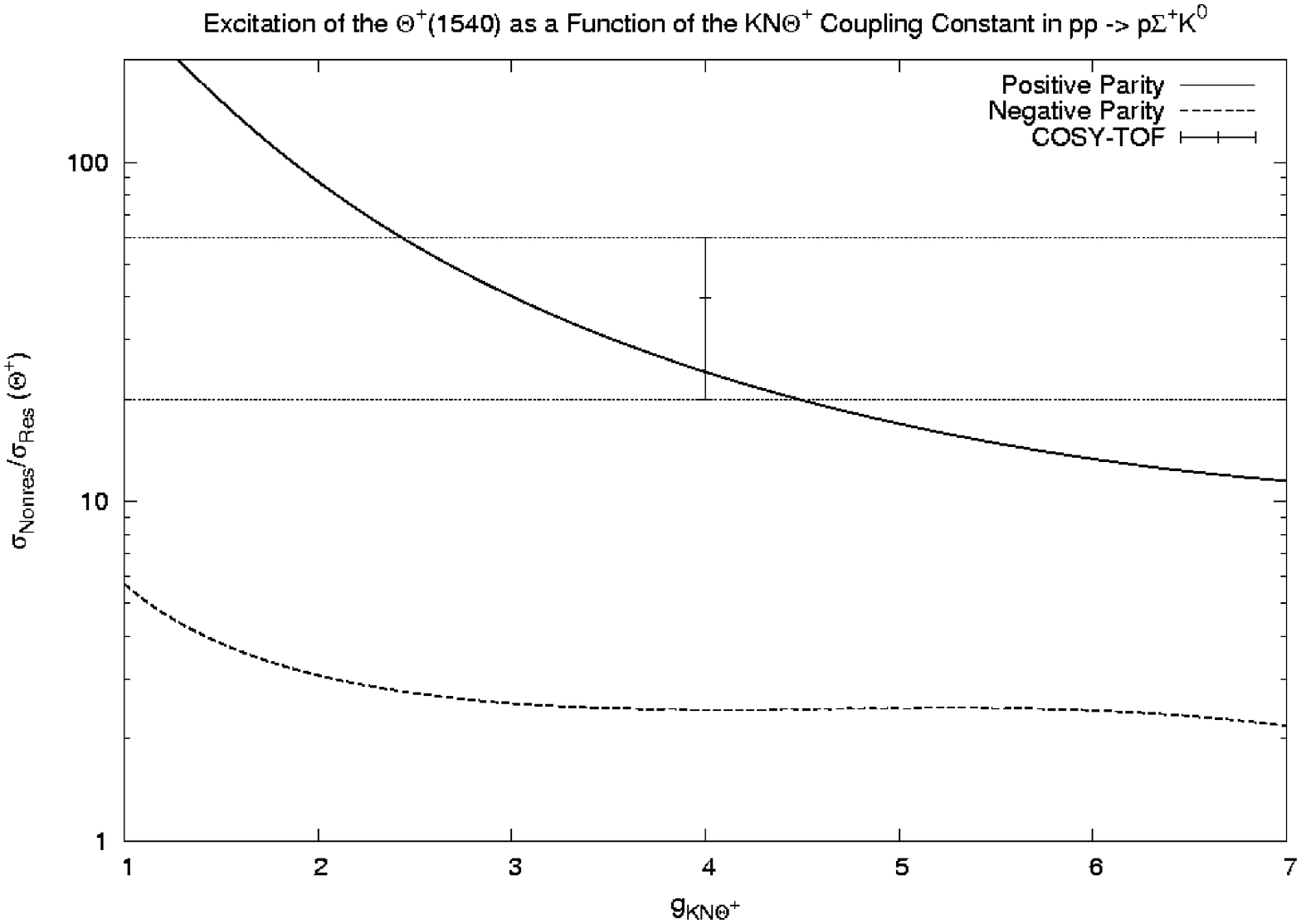}
\includegraphics[width=12.0cm,height=10.0cm,angle=0]{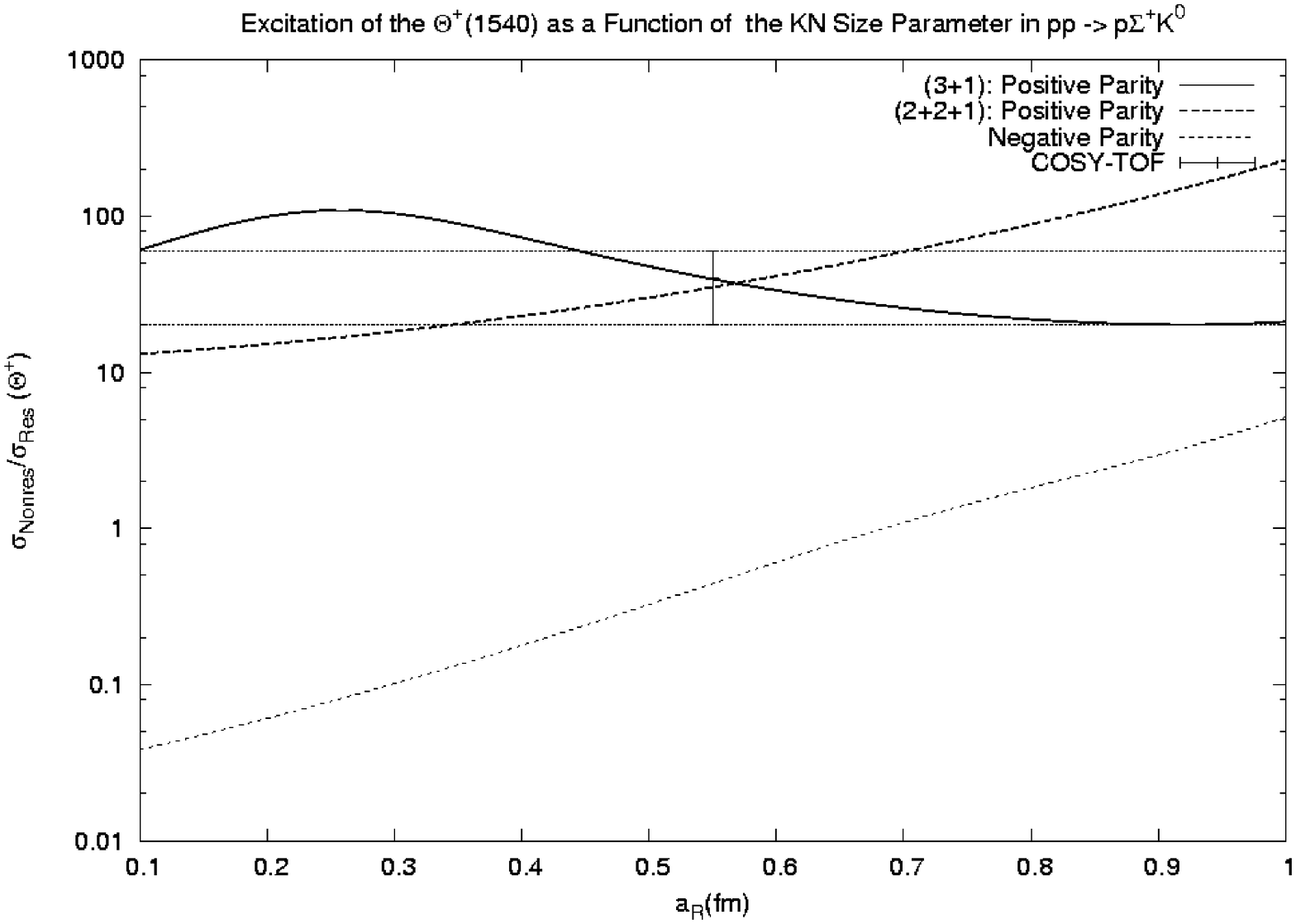}
\caption{Results for the ratio $\sigma_{NR}/ \sigma_R$ in a meson exchange (a) and a quark model calculation (b) for $\Theta^+$ with positive ($L_{KN}=1$) and negative ($L_{KN}=0$) parity. Shown in (a) is the dependence on the $N\Theta^+K$ coupling constant and in (b) for the width parameter $a_R$ of the KN cluster wave function (for (3+2) and (2+2+1) configurations of the $\Theta^+$). The experimental limits are tentatively indicated by the dotted lines}
\end{center}
\end{figure}

\begin{figure}[h!]
\begin{center}
\includegraphics[width=12.0cm,height=10.0cm,angle=0]{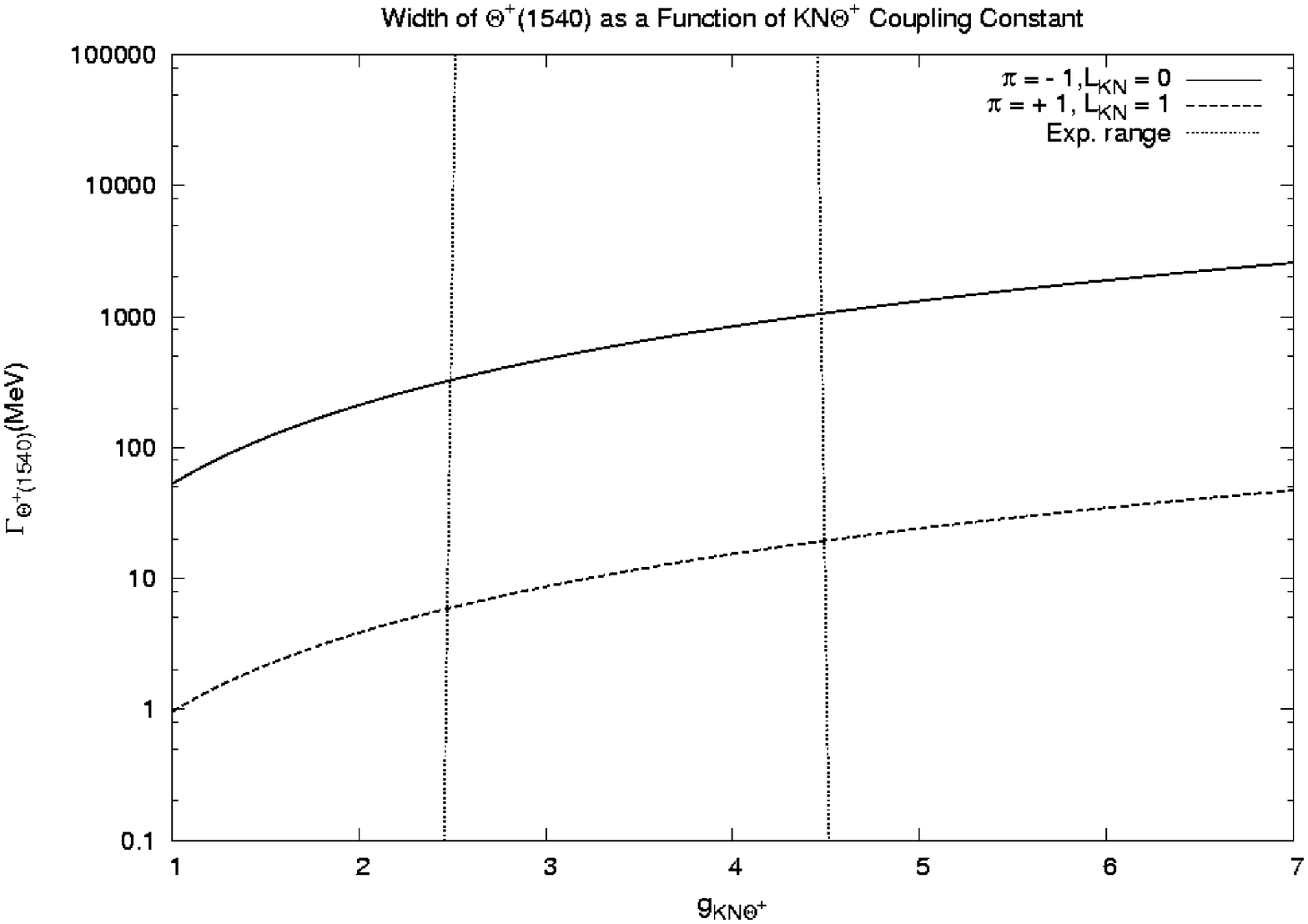}
\includegraphics[width=12.0cm,height=10.0cm,angle=0]{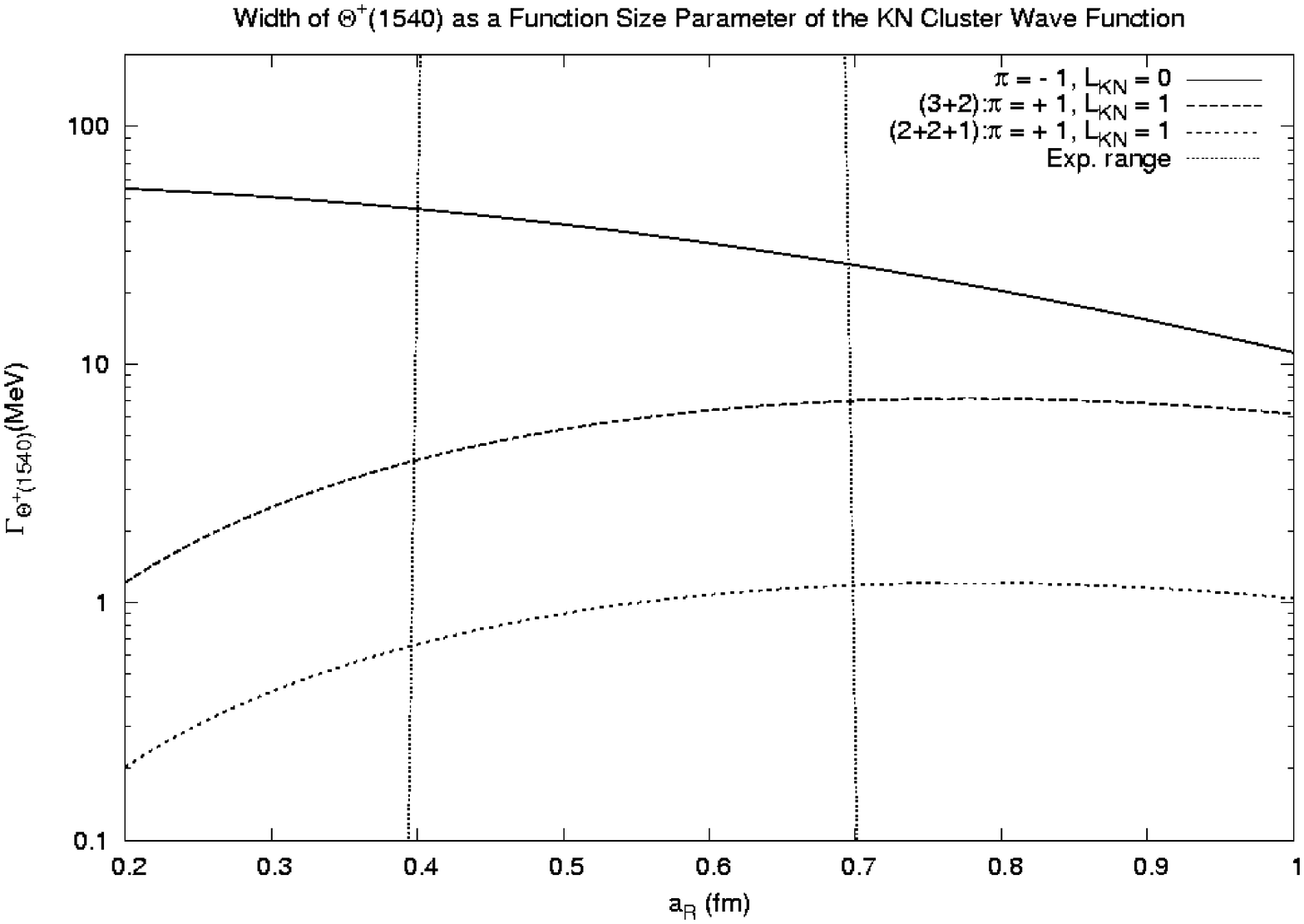}
\caption{As Fig. 4, however for the width $\Gamma_{\Theta^+}$ in a meson exchange (a) and a quark model calculation (b) for positive ($L_{KN} = 1$) and negative 
($L_{KN} = 0$) parity. The experimental range for $\Gamma_{\Theta^+}$ from Fig. 4 as a function of $g_{N\Theta^+K}$ and $a_R$ is indicated by the vertical lines}.
\end{center}
\end{figure}

\end{document}